\documentclass[%
 reprint,
 amsmath,amssymb,
 aps,
prl,
longbibliography
]{revtex4-1}

\usepackage[normalem]{ulem}
\usepackage{graphicx}
\usepackage{dcolumn}
\usepackage{bm}
\usepackage{bbm}
\usepackage{bbold}
\usepackage[dvipsnames]{xcolor}

\usepackage{hyperref}
\hypersetup{
    colorlinks=true,
    linkcolor=blue,
    filecolor=blue,      
    urlcolor=blue,
    citecolor=blue
}
\usepackage{float}

\begin{document}

\title{Resonant single-shot CNOT in remote double quantum dot spin qubits}

\author{Stephen R. McMillan}
\affiliation{Department of Physics, University of Konstanz, Konstanz D-78457, Germany}

\author{Guido Burkard}
\affiliation{Department of Physics, University of Konstanz, Konstanz D-78457, Germany}


\begin{abstract}
A critical element towards the realization of scalable quantum processors is non-local coupling between nodes. Scaling connectivity beyond nearest-neighbor interactions requires the implementation of a mediating interaction often termed a “quantum bus”. Cavity photons have long been used as a bus by the superconducting qubit community, but it has only recently been demonstrated that spin-based qubits in double quantum dot architectures can reach the strong coupling regime and exhibit spin-spin interactions via the exchange of real or virtual photons. Two-qubit gate operations are predicted in the dispersive regime where cavity loss plays a less prominent role. In this work we propose a framework for ac-driven quantum gates, in the context of a CNOT operation, between two non-local single-spin qubits dispersively coupled to a common mode of a superconducting resonator. We expect gate times near 150 ns and fidelities above 90\% with existing technology.
\end{abstract}

\maketitle

\textit{Introduction.} Electron spins in quantum dots (QD) have long been an attractive candidate for spin-based computation and information processing\cite{Loss1998, Hanson2007}. Spin initialization and readout has been demonstrated through spin-to-charge conversion\cite{Ono2002, Elzerman2004}, and remarkably long spin coherence times have been reached in Si based QDs as a result of isotopic purification\cite{Veldhorst2014}. Additionally, the use of local magnetic field gradients allows for all-electric spin manipulation\cite{Tokura2006, Pioro-Ladriere2008}. These milestones have lead to the recent realization of fault tolerant single and two-qubit gates \cite{Takeda2016, Yoneda2018, Mills2022, Xue2022, Noiri2022, Mills2022a}. With an eye on building quantum networks\cite{Fowler2012} and simulations\cite{Byrnes2008}, these single qubit systems can be scaled by an order of magnitude through tunnel-coupled linear and two-dimensional arrays which demonstrate spin coherent charge displacement\cite{Fujita2017,Flentje2017, Mukhopadhyay2018,Mortemousque2020}. These dense qubit arrays provide registers for implementaion of surface code protocols, but interconnection of these registers will require some form of long-range coupling\cite{Vandersypen2017}.

Achieving reliable long-range coupling in quantum systems is  perhaps the most pressing obstacle to realizing the next generation of quantum technology. Extending beyond micron-scale intermediate coupling schemes\cite{Hassler2015, Tosi2017} to macroscopic separations on the order of 1 mm requires coupling to an ancillary system\cite{Kimble2008,Schuetz2015}, a so-called ``quantum bus". Photonic coupling of superconducting qubits by way of superconducting resonators was realized well over a decade ago \cite{Majer2007} alongside proposals for coupling QD qubits with superconducting circuits\cite{Childress2004, Burkard2006, Srinivasa2016}. The relatively weak magnetic dipole coupling rate between a single photon and electron is inherently too slow (10-500 Hz) to overcome dephasing or cavity loss, presenting a clear challenge to reaching the strong spin-photon coupling regime\cite{Burkard2020}. Theoretical proposals were aimed a alleviating this issue by coupling the spin to the photon indirectly through the electric dipole moment\cite{Cottet2010}. Spin-charge hybridization along with the use of micromagnetics to establish local field gradients lead to the realization of strong spin-photon coupling for single spins in Si-based double QDs (DQD)\cite{Samkharadze2018, Mi2018}. 

Cavity mediated spin-spin interactions have been realized in both the resonant\cite{Borjans2020} and dispersive\cite{Harvey-Collard2022} regimes. Universal two-qubit photon mediated gates are predicted to exist for DQD spin qubits\cite{Benito2019a, Warren2019}, and driven two-qubit gates have been proposed for superconducting qubits, such as the so-called ``cross-resonance gate"\cite{Rigetti2010}. Despite these major advances, the possibilities for driving photon mediated entanglement of two spin qubits have been left unexplored. 

In this letter, we propose a framework for realizing a single-shot resonant CNOT gate between two single-electron DQD qubits dispersively coupled to a common cavity mode. By tuning the spin-charge hybridization these spin qubits are predicted to outperform their charge qubit counterpart\cite{Benito2019a}. Here we show that driving both spins at a common frequency and asymmetric amplitudes allows for controlled two-qubit rotations. We expect synchronization of the drive strength to produce a high-fidelity CNOT gate in roughly half the time required for the cross-resonance scheme. These gate times rival those predicted for local operations on driven exchange-coupled spin qubits\cite{Russ2018}. Given the sensitivity of QD spin qubits to local nuclei, gate fidelity should benefit from isotopically purified $^{28}$Si, where even in the presence of cavity loss and phonon emission fidelities larger than 90\% should be possible.

\begin{figure*}[ht!]
\begin{centering}
       \includegraphics[width=\textwidth,scale=0.5]{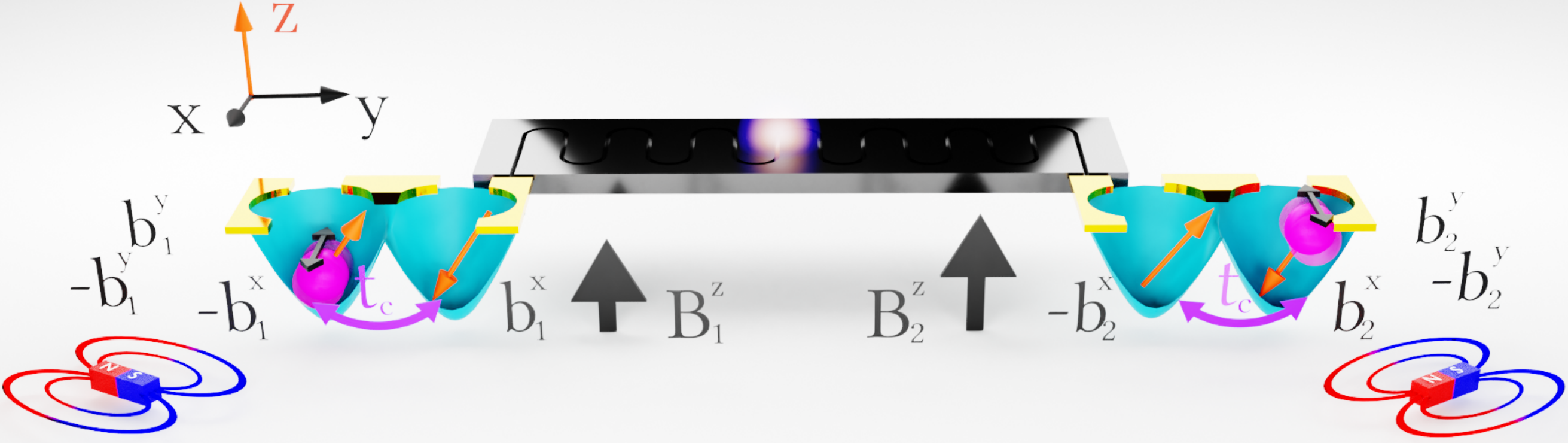}
       \caption[width=\columnwidth]
{Diagram of the system. Two DQDs are coupled to the same cavity mode of a microwave resonator. Local gradients $b^x_j$ provide an artificial spin-orbit coupling, allowing the spin to couple to the photon through the electric dipole of the DQD. A gradient in the external field, $\delta B^z=B^z_2-B^z_1$, lifts the degeneracy between anti-parallel spin configurations allowing the individual spins to be addressed. An additional local gradient, $b^y$, provides an electrically driven ac magnetic field which is used to drive spin transitions.}\label{fig:Setup}
\end{centering}
\end{figure*}

\textit{Model.}
We consider two singly-charged DQDs capacitively coupled to a common mode of a superconducting microwave resonator (cavity), with frequency $\omega_c$. The energy detuning between the right (R) and left (L) dots, $\epsilon$, as well as the tunnel coupling, $t_c$, are electrically controlled, and the spin-photon interaction is facilitated by a local inhomogenous magnetic field $b^x$. We begin in the double-dot basis and consider a model Hamiltonian $\tilde{H} = \tilde{H}_0+\tilde{H}_d(t)+\tilde{H}_I$, with $\tilde{H}_0 = \sum_j ( t_c \tilde{\tau}^x_j +\epsilon_j\tilde{\tau}_z/2 + B^z_j \tilde{\sigma}^z_j/2 + b^x \tilde{\sigma}^x_j/2) + \omega_c a^\dagger a$. In this expression, $B^z_j$ is the external field, $a^\dagger$ $(a)$ is the photonic creation (annihilation) operator, $\tilde{\sigma}^k_j$ for $k=\{x, y, z\}$, are the spin Pauli matrices, and $\tilde{\tau}^k_j$ for $k=\{x, y, z\}$, are the spatial Pauli matrices satisfying the relation $\tau^z_j|L_j(R_j)\rangle=~(-1)|L_j(R_j)\rangle$. The driving Hamiltonian is written as $\tilde{H}_\text{d}(t) = \sum_j b^y_{j} \cos (\omega^d_j t + \phi) \tilde{\sigma}^y_{j}/2$, where $b^y_j$ is the driving strength, and the driving frequency, $\omega^d_j$, can be controlled electrically by shaking the spin in an inhomogeneous magnetic field - electrically driven spin resonance (EDSR)\cite{Tokura2006, Pioro-Ladriere2008, Kawakami2014, Zajac2018, Mi2018, Samkharadze2018}.
The charge couples to the cavity mode through a dipolar interaction with the Hamiltonian $\tilde{H}_\text{int} = \sum_j g^c  (a^\dagger + a)\tilde{\tau}^z_j$. The electric dipole coupling strength $g_c$ is proportional to the dipole moment of the DQD, and is therefore electrically controllable through $\epsilon$. In what follows we maximize the dipole moment of the DQD by choosing the symmetric configuration where the charge is equally likely to be in the R or L dot ($\epsilon=0$).

The tunnel coupling $t_c$ hybridizes the double-dot states to form the symmetric and antisymmetric states $|\pm_j\rangle=(|L_j\rangle\pm|R_j\rangle)/\sqrt{2}$. These states hybridize with the spin due to the local gradient $b^x$. Recasting the Hamiltonian in the basis of $\tilde{H}_0$ leads to the transformed components,
\begin{equation}
\begin{aligned}
    H_0 &=\frac{1}{2}\sum_j (E^\tau_j\tau^z_j + E^\sigma_j\sigma^z_j) + \omega_c a^\dagger a,\\
    H_\text{d}(t) &= \frac{1}{2}\sum_j (D^\tau_j \tau^y_j\sigma^z_j+D^\sigma_j\sigma_j^y) \cos(\omega^d_j t+\phi),\\
    H_I &= \sum_j (g^\tau_j \tau^x_j- g^\sigma_j\sigma^x_j\tau^z_j)(a^\dagger + a),
\end{aligned}
\end{equation}
where the driving strengths are $D^{\sigma}_j= b^y_j \cos\bar{\theta}_j$ and $D^{\tau}_j=b^{y}_j \sin\bar{\theta}_j$, and the cavity coupling strengths are $g^\sigma_j=g_c\sin\bar{\theta}_j$ and $g^\tau_j=g_c\cos\bar{\theta}_j$ for spin and charge respectively. These terms depend on the average hybridization angle $\bar{\theta}_j=(\theta^+_j+\theta^-_j)/2$, where $\theta^\pm_j=\arctan [b^x/(2t_c \pm B^z_j)] \in [0,\pi]$. In what follows we assume $2t_c > B^z_j \gg b^x$, such that in analogy to Ref.~\cite{Benito2019a},
the charge (spin) transition energies are given by $E^{\tau(\sigma)}_j = E_{2(1),j}-E_{0,j}$ where
$E_{2(1),j} = \pm\sqrt{(2t_c-B^z_j)^2+(b^x)^ 2)}/2$ and
$E_{0,j} =~{ -\sqrt{(2t_c+B^z_j)^2+(b^x)^2)}/2}$.

Dispersive coupling between the DQDs and the cavity mitigates losses due to interactions with real photons, limiting the spin-photon exchange to short-lived virtual transitions \cite{Harvey-Collard2022, Bonsen2022}. The dispersive regime requires the cavity frequency to be detuned from the spin transition energy by an amount $\Delta_j=E^\sigma_j-\omega_c$ such that $\Delta_j\gg\hbar g^\sigma_j$. In this regime the photonic states are weakly coupled, and we perform a Schrieffer-Wolff transformation to decouple the vacuum state from the populated photon states. Tuning the cavity frequency such that $g_j^\tau\Delta_j/|E_j^\tau-\omega_c|\ll g_j^\sigma\ll\Delta_j$ is most interesting for spin-qubit operation\cite{Benito2019a}. In the rotating wave approximation the dispersive Hamiltonian can be written as
\begin{equation}
\label{eq:TD_effH}
\begin{aligned}
    \mathcal{H} = &\sum_j\bigg[\tilde{E}^\sigma_\text{avg} \sigma^z_j 
    + i D^\sigma_j\bigg(e^{i(\omega_j^d t +\phi_j)}\sigma^-_j-e^{-i(\omega_j^d t +\phi_j)}\sigma^+_j\bigg)/4\bigg]\\
    &+\frac{1}{2}\delta\tilde{E}^\sigma (\sigma^z_1 - \sigma^z_2)
    + \frac{\mathcal{J}}{2} (\sigma^+_1\sigma^-_2+\sigma^-_1\sigma^+_2),
\end{aligned}
\end{equation}
where the Pauli matrices $\sigma^k_j$ and $\tau^k_j$ now correspond to the qubits dressed by the photonic excitations, and the cavity mediated spin-spin interaction coefficient is $\mathcal{J}=~g^\sigma_1 g^\sigma_2(\frac{1}{\Delta_1}+\frac{1}{\Delta_2})$. Due to the difference in external field between the two qubits, the dressed energy levels, $\tilde{E}^\sigma_j = \frac{1}{2}[E^\sigma_j + 2 (g^\sigma_j)^2\frac{E^\sigma_j}{(E^\sigma_j)^2-\omega_c^2}]$, have been written in terms of an average spin-flip energy $\tilde{E}^\sigma_\text{avg} = (\tilde{E}^\sigma_1 + \tilde{E}^\sigma_2)/2$ and a difference term $\delta \tilde{E}^\sigma ~=~ \tilde{E}^\sigma_1 - \tilde{E}^\sigma_2 $. Since the cavity is tuned closer to the spin transition energy, any non-local charge-charge and spin-charge transitions are highly off-resonant and therefore provide a negligible contribution to the dynamics. Additionally, by tuning the local driving to spin transitions the local charge transitions are negligible for the same reason, and the resulting Hamiltonian has no relevant dynamics in the charge channel. 

We now assume that both spins are driven at the same frequency ($\omega^d_1=\omega^d_2=\omega^d$), but not necessarily the same strength. Next we assume the spin-spin interaction to be small relative to the spin-flip gradient, $\mathcal{J}\ll\delta\tilde{E}^\sigma$, and use the adiabatic basis, $\{ |\uparrow\uparrow\rangle,|\widetilde{\downarrow\uparrow}\rangle,|\widetilde{\uparrow\downarrow}\rangle,|\downarrow\downarrow\rangle\}$, where $\mathcal{H}$ is diagonalized with respect to the spin-spin interaction. In this basis the only off-diagonal contribution comes from the driving term, which can be written as
\begin{equation}
\begin{aligned}
   \hat{\mathcal{H}}_\text{d}(t)= &\frac{1}{2}D^\sigma_1\cos(\omega^d t+\phi)\bigg[\hat{\sigma}^y_1+\frac{\mathcal{J}}{2\delta\tilde{E}^\sigma}\hat{\sigma}^z_1\hat{\sigma}^y_2\bigg]\\
    &+\frac{1}{2}D^\sigma_2\cos(\omega^d t+\phi)\bigg[\hat{\sigma}^y_2 - \frac{\mathcal{J}}{2\delta\tilde{E}^\sigma}\hat{\sigma}^y_1\hat{\sigma}^z_2\bigg].
\end{aligned}
\label{eq:Hdriving}
\end{equation}

\begin{figure}[t]
\begin{centering}
       \includegraphics[width=0.8\columnwidth]{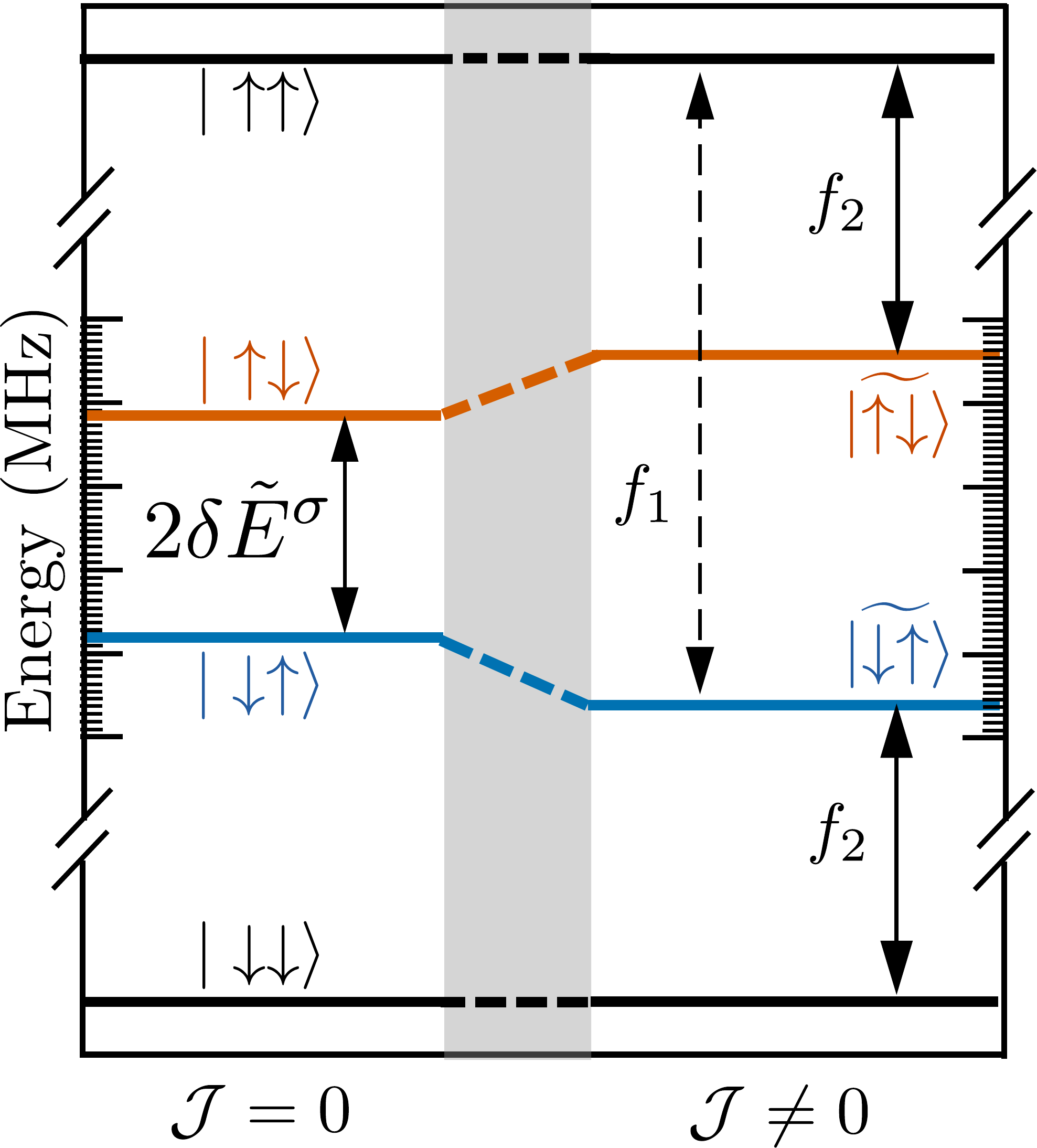}
       \caption[width=\columnwidth]
{Sketch of the energy levels for the spin manifold in the lowest charge sector. A gradient in the external field lifts the degeneracy between the $m_s=0$ states. Once the cavity mediated spin-spin interaction is present the $m_s=0$ levels experience an additional symmetric shift in frequency. Frequency $f_1$($f_2$) drives the spin at DQD1(2) unconditionally due to the symmetry of the system.}\label{fig:EnergyLevels}
\end{centering}
\end{figure}

Figure~\ref{fig:EnergyLevels} shows a sketch of the energy levels for states in the lowest charge sector. The effective gradient in the external field, $\delta \tilde{E}^\sigma$, lifts the degeneracy between the $m_s=0$ states. The presence of the non-local spin-spin coupling adds an additional symmetric frequency shift to the $m_s=0$ states in the adiabatic basis in contrast to asymmetric shift seen for homogeneous exchange coupling\cite{Russ2018}. The symmetry of the system dictates that frequency $f_1$($f_2$) drives the spin at DQD1(2) unconditionally.

To eliminate the time dependence in Eq.~(\ref{eq:Hdriving}), we work in  a frame rotating at the drive frequency about the spin quantization axis. The rotating frame Hamiltonian has the form $\mathring{\mathcal{H}}=~i \dot{\hat{R}}(t)\hat{R}^\dagger(t) + \hat{R}(t)\hat{\mathcal{H}}(t)\hat{R}^\dagger(t)$, where $\hat{R}=~\exp[i \omega^d(\hat{\sigma}^z_1+\hat{\sigma}^z_2)t/2]$. The time-independent Hamiltonian can be expressed as
\begin{equation}\label{eqn:CNOT_Ham}
\mathring{\mathcal{H}}=
\begin{pmatrix}
\lambda_\parallel & -i \alpha_-^* & -i \beta_+^* & 0\\
i \alpha_- & -\lambda_\perp & 0 & -i\beta_-^*\\
i\beta_+ & 0 & \lambda_\perp & -i\alpha_+^*\\
0 & i\beta_- & i\alpha_+ &  -\lambda_\parallel\\
\end{pmatrix},
\end{equation}
with $\lambda_\parallel=2 \tilde{E}^\sigma_\text{avg}-\omega^d$, $\lambda_\perp = \delta \tilde{E}^{\sigma}+\frac{\mathcal{J}^2}{2\delta \tilde{E}^{\sigma}}$, and the effective driving amplitudes
\begin{equation}
    \alpha_\pm \approx \frac{1}{4}\bigg[\pm D^{\sigma}_{2}\frac{\mathcal{J}}{2\delta\tilde{E}^\sigma}+ D^{\sigma}_{1}\bigg]e^{i\phi}
\end{equation}
and
\begin{equation}
    \beta_\pm \approx \frac{1}{4}\bigg[\pm D^{\sigma}_{1}\frac{\mathcal{J}}{2\delta\tilde{E}^\sigma}+ D^{\sigma}_{2} \bigg]e^{i\phi}.
\end{equation}
In what follows we rely on the hybridized driving amplitudes to generate single-shot remote entanglement.

 \textit{Resonant non-local {\rm CNOT} gate.} One route to entangling two spins is to perform a gate with conditional rotations. For a CNOT gate we define the spin in DQD1 to be the \textit{target} and the spin in DQD2 to be the \textit{control}. Using the basis above, this operation is represented by the matrix
 \begin{equation}\label{eq:CNOTmat}
     U_\text{CNOT} = 
     \begin{pmatrix}
      0 & 1 & 0 & 0\\
      1 & 0 & 0 & 0\\
      0 & 0 & 1 & 0\\
      0 & 0 & 0 & 1
     \end{pmatrix}.
 \end{equation}
 To model this operation we require the target spin to flip if the control is spin-up and an identity operation if the control is spin-down.
 To resonantly drive the spin at DQD1 we choose a driving frequency that is determined by the energy separation between $|\uparrow\uparrow\rangle$ and $|\widetilde{\downarrow\uparrow}\rangle$:
 $\omega^d-\delta\omega^d=\lambda_\parallel + \lambda_\perp$. On resonance, $\delta\omega^d=0$, the $\beta$ transitions couple states split in energy by the relatively large quantity $2 \delta \tilde{E}^{\sigma}$. As long as the off-resonant transitions are weakly coupled ($\beta_\pm \ll \delta\tilde{E}^\sigma$) the Hamiltonian expressed in  Eq.~(\ref{eqn:CNOT_Ham}) is effectively block diagonal, with both $\{|\uparrow\uparrow\rangle,|\widetilde{\downarrow\uparrow}\rangle\}$ and $\{|\widetilde{\uparrow\downarrow}\rangle,|\downarrow\downarrow\rangle\}$ blocks producing full Rabi oscillations at frequencies $\Omega_-=2|\alpha_-|$ and $\Omega_+=2|\alpha_+|$ respectively. The time required for maximum population transfer between the states $|\uparrow\uparrow\rangle\leftrightarrow|\downarrow\uparrow\rangle$ is $t_\text{CNOT}=(2m+1)\pi/\Omega_-$, for integer $m$.


Treating the resonant blocks individually we calculate the unitary operator for each. With the aim of obtaining the unitary in Eq.~(\ref{eq:CNOTmat}), the drive phase should be set such that the rotation is about the $\pm$x-axis: $\phi=(2\ell+1)\pi/2$, for integer $\ell$. Under this condition, the control spin-up block evolves as
\begin{equation}\label{eqn:CNOT_unitary2up}
    U_{2=\uparrow}=e^{-i f_1 t}
    \bigg[ \cos\bigg(\frac{\Omega_-t}{2}\bigg)\mathbb{1} + i^{2\ell+1} \sin\bigg(\frac{\Omega_-t}{2}\bigg) \sigma^x\bigg],
\end{equation}
and the block for control spin down evolves as
\begin{equation}\label{eqn:CNOT_unitary2down}
    U_{2=\downarrow}=e^{-i f_2 t}
    \bigg[ \cos\bigg(\frac{\Omega_+t}{2}\bigg)\mathbb{1} + i^{2\ell+1} \sin\bigg(\frac{\Omega_+t}{2}\bigg) \sigma^x\bigg].
\end{equation}
The dynamic phase is dependent on the energy of the corresponding block, $f_1 = \delta \tilde{E}^\sigma + \mathcal{J}^2/2\delta \tilde{E}^\sigma$ and $f_2 =- \delta \tilde{E}^\sigma - \mathcal{J}^2/2\delta \tilde{E}^\sigma$. The choice of phase allows one to write the total unitary ($U=U_{2=\downarrow}\oplus U_{2=\uparrow}$) in the form of the CNOT unitary shown in Eq.~(\ref{eq:CNOTmat}), up to local rotations about the z-axis. However, one could generate a conditional rotation about any arbitrary axis in the x-y plane by making the substitution $\sigma^x\to\sin\phi\;\sigma^x+\cos\phi\;\sigma^y$.

Under a symmetric drive ($b^y_1=b^y_2$) the difference in Rabi frequency between the upper and lower block differs only by the small quantity $\mathcal{J}/\delta\tilde{E}^\sigma$, leading to indiscriminate rotation of the target spin. In order to generate a single-shot conditional rotation of the target spin, the Rabi frequencies $\Omega_+$ and $\Omega_-$ should be out of phase. The necessary condition is then
\begin{equation}\label{eqn:CNOT_RabiSync}
    \Omega_-=\frac{2m+1}{2n}\Omega_+,
\end{equation}
with integer $n$. This can be done by adjusting the ratio of the ac driving strengths such that
\begin{equation}\label{eqn:SyncCond}
    \frac{D^\sigma_{1}}{D^\sigma_{2}}=-\frac{1+2(m+n)}{1+2(m-n)}\frac{\mathcal{J}}{2 \delta \tilde{E}^\sigma_z},
\end{equation}
where $m$ and $n$ obey Eq.~(\ref{eqn:CNOT_RabiSync}). The unitary operator up to a global phase at time $t_\text{CNOT}$ is
\begin{equation}
    U(t_\text{CNOT})=e^{i(\Phi_\text{dyn} + \Phi_\text{hol}) \sigma^z_2}U_\text{CNOT}.
\end{equation}
The time dependent dynamic phase is $\Phi_\text{dyn}=\frac{f_2-f_1}{2}t_\text{CNOT}$ and the time-independent holonomic phase is $\Phi_\text{hol}=\frac{\pi}{2}[(2\ell+1)/2+m-n]$.

\begin{figure}[t]
       \includegraphics[width=\columnwidth,scale=0.25]{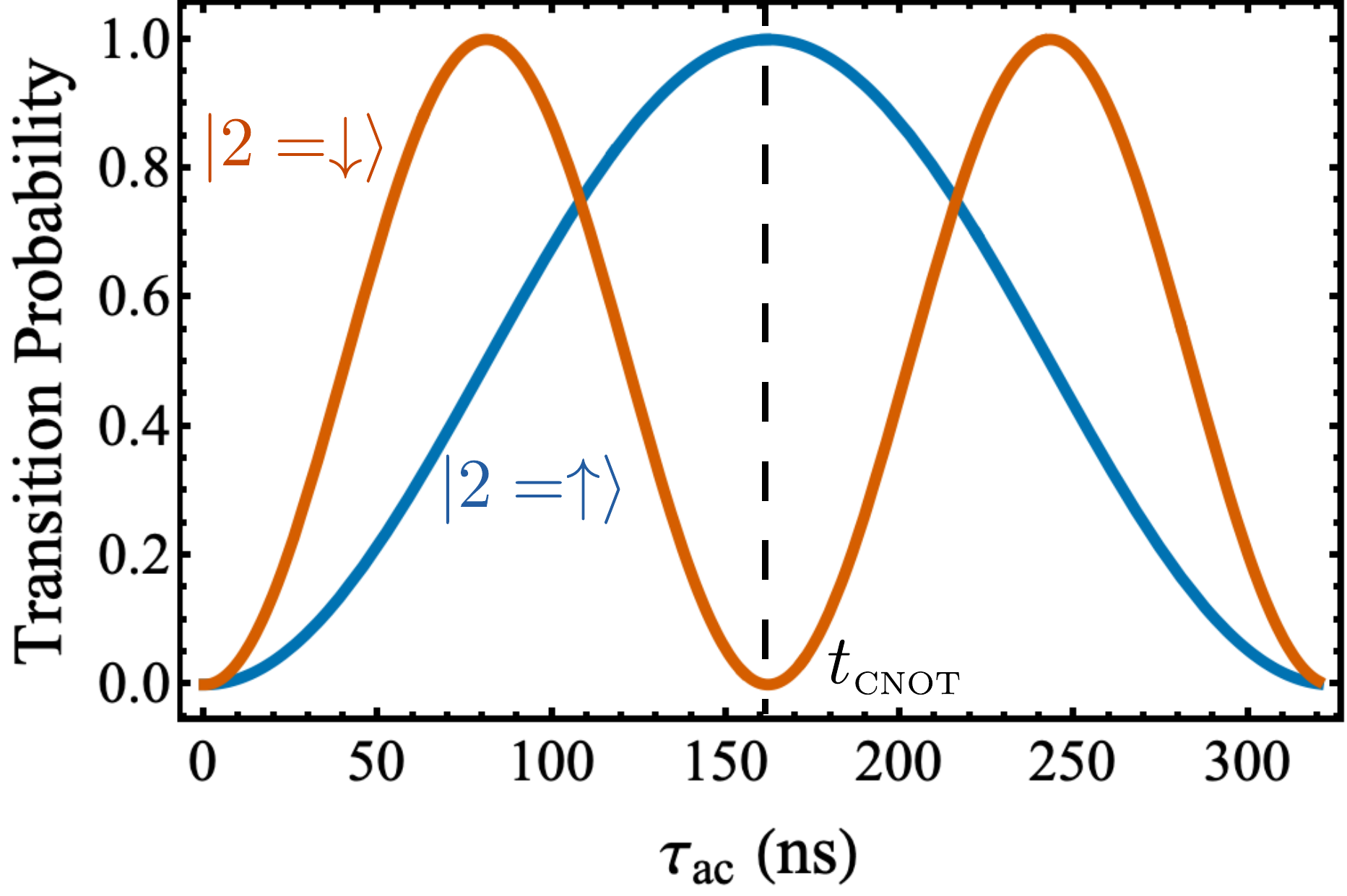}
       \caption[width=\columnwidth]
{Calculated transition probabilities for the synchronized $|2=\uparrow\rangle$ (blue) and $|2=\downarrow\rangle$ (red) blocks. The relevant parameters are chosen to be experimentally realizable with a driving strength $D^\sigma_{2}/h=100$ MHz. The integers $m=0$ and $n=1$ are chosen to minimize the gate time, shown here to be $t_\text{CNOT}=162$ ns. Parameters: $t_c/2\pi$=3.9 GHz, $B^z_\text{avg}/2\pi$=6.3 GHz, $\delta B^z/2\pi$=300 MHz, $b^x/2\pi$=780 MHz, $\omega_c/2\pi$=5.6 GHz, $g_c/2\pi$=190 MHz.}\label{fig:CNOT}
\end{figure}

Figure~\ref{fig:CNOT} shows the synchronized Rabi oscillations for the driven CNOT gate. System parameters were chosen to represent realistic experimental conditions and are quoted in the figure caption. The analytic expressions derived above for the unitary evolution ignore the off-resonant transitions with coupling $\beta_\pm$ in Eq.~(\ref{eqn:CNOT_Ham}). In general, strong driving will increase population transfer through the off-resonant channels impacting the coherent average gate fidelity $\bar{F}_c$. Numerical calculations of this fidelity show that for a modest synchronized drive strength of $D^\sigma_2=100$ MHz we predict gate times of $t_\text{CNOT}=162$ ns with coherent average fidelity $\bar{F}_c=0.99$. In principle, driving with this strength requires a field gradient with magnitude near 3.5 mT, a requirement well within the typical strength of contemporary micromagnetics ($\sim$1 mT/nm).

\textit{Discussion and summary.} In addition to the parasitic effects of off-resonant channels, coupling to the environment also reduces the fidelity of the gate. Since the spin qubit couples to the cavity through a virtual charge qubit, the gate is vulnerable to both electron-phonon interaction as well as photonic loss through the cavity. However, since the charge qubit and photons are both virtual the influence on gate fidelity is less pronounced than for resonant coupling. To estimate these effects we calculate the average fidelity\cite{Nielsen2002} in the presence of cavity loss, electron-phonon coupling, and general spin decoherence. The master equation in the dispersive basis then has the form
\begin{equation}\label{eqn:MastEqn}
\begin{aligned}
    \frac{d}{dt}\hat{\rho}^\sigma(t) = &-i [\hat{\mathcal{H}}(t),\hat{\rho}^\sigma (t)] +\frac{\kappa}{2}\sum_j\frac{g^\sigma_j}{\Delta_j}\mathcal{D}[\hat{\sigma}^-_j]\hat{\rho}^\sigma (t)\\
    &+\frac{\gamma_p}{2}\sum_j\sin(\bar{\theta}_j)\mathcal{D}[\hat{\sigma}^-_j]\hat{\rho}^\sigma (t)\\ 
    &+ \frac{\nu_{T^*_2}}{2}\sum_j\mathcal{D}[\hat{\sigma}_j^z]\hat{\rho}^\sigma (t),
\end{aligned}    
\end{equation}
where $\mathcal{D}[\hat{o}]$ represents the Lindblad superoperator $\mathcal{D}[\hat{o}]\rho=2\hat{o}\rho \hat{o}^\dagger+\{\hat{o}^\dagger \hat{o},\rho\}$, and $\{...\}$ represents the anti-commutator. The density matrix $\hat{\rho}^\sigma$ corresponds to the two-spin subspace in the dispersive adiabatic basis, and we assume the environment to be at zero temperature. The fist term describes coherent evolution of the system, the second term describes loss through the cavity at rate $\kappa$, the third term accounts for the relaxation due to electron-phonon coupling at rate $\gamma_p$, and the last term expresses general dephasing of the spin at rate $\nu_{T^*_2}$. 

An average measure of how well a trace-preserving quantum operation, $\mathcal{E}$, approximates a quantum gate, $U$, is defined as  $\bar{F}(\mathcal{E},U)\equiv \int d\psi \langle\psi|U^\dagger\mathcal{E}(\psi)U|\psi\rangle$. Making use of the entanglement fidelity\cite{Schumacher1996}, $F_e$, the average fidelity in a four-dimensional Hilbert space can be written in a compact form\cite{Nielsen2002}: $\bar{F}(\mathcal{E},U) = (\sum_j \text{Tr}[UU^\dagger_jU^\dagger\mathcal{E}(U_j)]+16)/80$. Here we introduce a basis of unitary operators, $U_j = X^k Z^l$, on the four-dimensional two-qubit subspace, where $X|j\rangle = |j\oplus 1\rangle$ and $Z|j\rangle = e^{2\pi i j /4}|j\rangle$ are defined on computational basis states $|0\rangle,...,|3\rangle$. Here the operation $\oplus$ represents addition modulo 4.

\begin{figure}[t!]
       \includegraphics[width=\columnwidth]{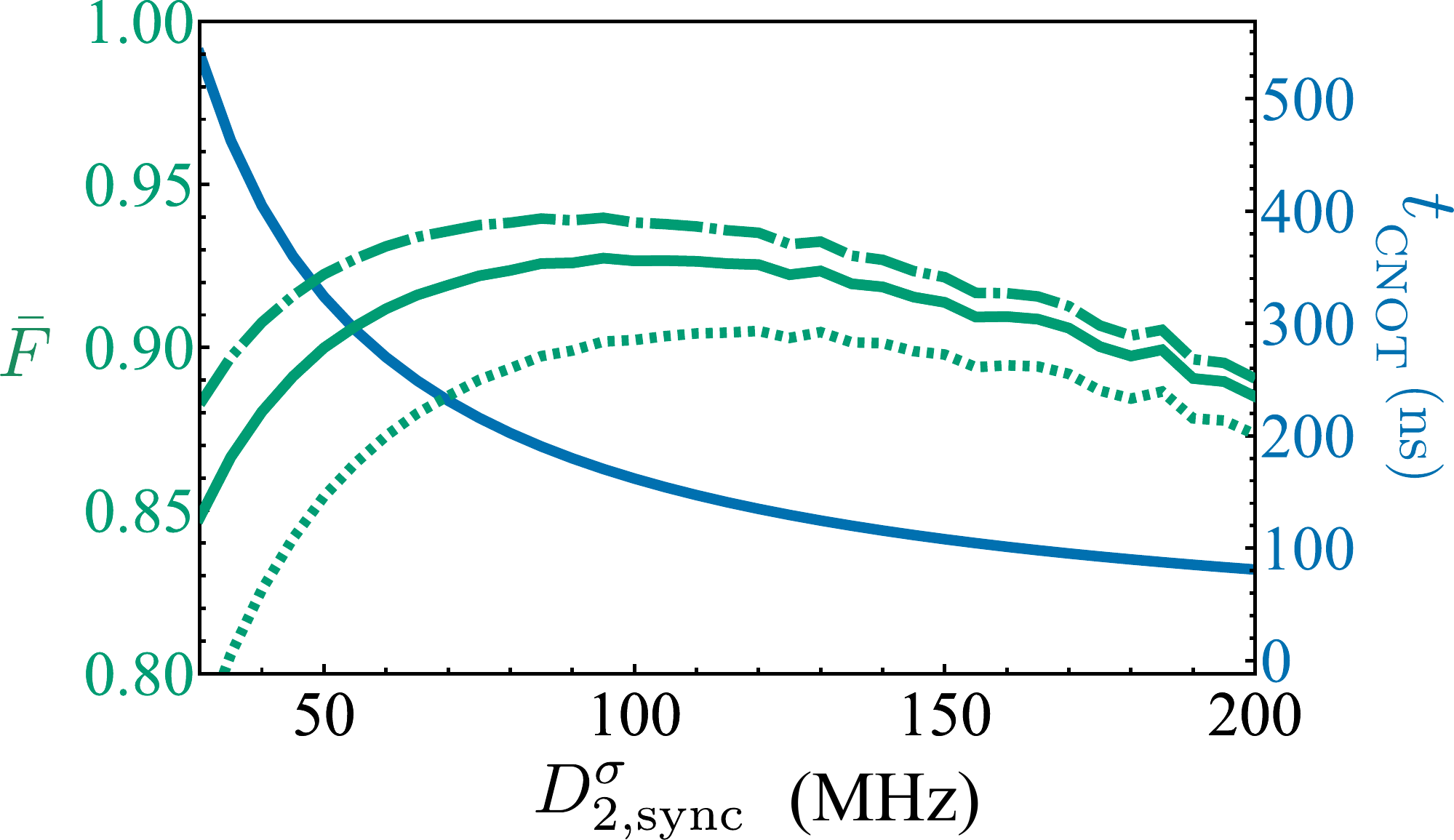}
       \caption[width=\columnwidth]
{The gate time and average fidelity as a function of synchronized drive strength for $\nu_{T^*_2}=100$ kHz with $\gamma_p=4$ MHz (green dotted) and $\gamma_p=1$ MHz (green solid), and $\nu_{T^*_2}=50$ kHz for $\gamma_p=1$ (green dot-dashed).   Additional parameters: $t_c/2\pi$=3.9 GHz, $B^z_\text{avg}/2\pi$=6.3 GHz, $\delta B^z/2\pi$=300 MHz, $b^x/2\pi$=780 MHz, $\omega_c/2\pi$=5.6 GHz, $g_c/2\pi$=190 MHz, and $\kappa=1.5$ MHz.}\label{fig:FvsTvsD}
\end{figure}

The average fidelity and gate time are plotted in Fig.~\ref{fig:FvsTvsD} as a function of the synchronized drive strength $D^\sigma_{2,\text{sync}}$ for three sets of realistic parameters. Strong driving leads to faster gate times (shown by the blue line), but the fidelity is limited by increased transitions through off-resonant channels. However, weaker driving requires the system to be highly coherent, and if the gate time is too long incoherent processes become significant and the fidelity again suffers. Competition between off-resonant evolution and dissipation results in an optimal drive strength that maximizes gate fidelity. The optimal synchronized drive strength for the set of parameters shown in Fig.~(\ref{fig:FvsTvsD}) are within the range $D^\sigma_{2,\text{sync}}\approx 95-120$ MHz. The top-end parameters yield an average fidelity of $\bar{F}=0.94$, as shown by the green dot-dashed line. However, using rather conservative values \cite{Sigillito2019, Stano2021} still yields an average fidelity above 90\% ($\bar{F}=0.905$) as shown by the dotted green line.  


With recently achieved electric dipole coupling strengths $g_c/2\pi\approx 190$ MHz using high impedance resonators\cite{Harvey-Collard2022}, we predict gate fidelities well above 90\% in the presence of cavity loss, phonon emission, and spin dephasing. Gate times around 150 ns are relatively fast compared to other non-local entanglement operations, such as the cross-resonance gate\cite{Rigetti2010} which is typically 300-400 ns. Along with single-spin operations this work elucidates the potential for all electrical remote entanglement in solid-state DQD spin qubits, and enhances the prospects of these systems for quantum information processing.

\begin{acknowledgments}
We would like to acknowledge J. Mielke for useful and valuable discussions.
\end{acknowledgments}

\bibliography{central-refs.bib}

\end{document}